\documentclass[aps,twocolumn,prd,amsmath,nofootinbib,preprintnumbers,superscriptaddress]{revtex4-1}
\usepackage{epsfig,slashed,nicefrac}
\usepackage[utf8]{inputenc}
\usepackage{color}
\usepackage{float} 
\usepackage{subfig} 
\usepackage{eqnarray,amsmath}
\usepackage{multirow}

\newcommand{\harpoon}{\overset{\rightharpoonup}}

\def\be{\begin{equation*}}
\def\ee{\end{equation*}}
\def\bsp#1\esp{\begin{split}#1\end{split}} 
\def\bpm{\begin{pmatrix}}
\def\epm{\end{pmatrix}}

\begin{document}


\title{Probing impact-parameter dependent nuclear parton densities from double parton scatterings in heavy-ion collisions}

\author{Hua-Sheng Shao}
\affiliation{Laboratoire de Physique Th\'eorique et Hautes Energies (LPTHE), UMR 7589, Sorbonne Universit\'e et CNRS, 4 place Jussieu, 75252 Paris Cedex 05, France}

\date{\today}

\begin{abstract}
We propose a new method to determine the spatially or impact-parameter dependent nuclear parton distribution functions (nPDFs) using the double parton scattering (DPS) processes in high-energy heavy-ion (proton-nucleus and nucleus-nucleus) collisions. We derive a simple generic DPS formula in nuclear collisions by accommodating both the nuclear collision geometry and the spatially dependent nuclear modification effect, under the assumption that the impact-parameter dependence of nPDFs is only related to the nuclear thickness function. While the geometric effect is widely adopted, the impact of the spatially dependent nuclear modification on DPS cross sections has been overlooked so far, which can, however, be significant when the initial nuclear modification is large. In turn, the DPS cross sections in heavy-ion collisions can provide useful information on the spatial dependence of nPDFs. They can be, in general, obtained in minimum-bias nuclear collisions, featuring the virtue of independence of Glauber modeling. 
\end{abstract}

\maketitle


\section{Introduction}

Multiple particle production at high-energy hadron colliders, such as at the LHC, is dominated by simultaneous multiple interactions between partons from the initial hadrons. Such multiple-parton interactions (MPIs) are indispensable in scrutinizing many event activities and hadron multiplicities at colliders. One particularly interesting case is that, when more-than-one reactions in a collision are lying at hard scales, the perturbative QCD approach based on the factorization theorem~\cite{Diehl:2018wfy} or conjecture applies. The studies of the so-called hard multiple-parton scattering processes can deepen our understanding of QCD and the possible multiple-parton correlations in a nucleon (see, e.g., Refs.~\cite{Blok:2013bpa,Kasemets:2014yna,Rinaldi:2016jvu,Rinaldi:2018bsf,Rinaldi:2018slz}). They provide new means to access the information of the nonperturbative structure of hadrons, which is complementary to the one obtained from nucleon one-body distributions. Because of the power counting of the cross sections, the leading multiple-parton scattering mechanism is the double parton scattering (DPS), where only two partonic scattering subprocesses happen at the same time.

In the LHC era, we have witnessed rapid theoretical developments~\cite{Blok:2011bu,Diehl:2011yj,Diehl:2011tt,Gaunt:2011xd,Ryskin:2011kk,Snigirev:2010ds,Ceccopieri:2010kg,
Manohar:2012jr,Chang:2012nw,Manohar:2012pe,
Gaunt:2012dd,Rinaldi:2013vpa,Blok:2013bpa,Diehl:2014vaa,Golec-Biernat:2014bva,Rinaldi:2014ddl,Kasemets:2014yna,Diehl:2015bca,Rinaldi:2016jvu,Buffing:2017mqm,
Diehl:2017kgu,Vladimirov:2017ksc,Diehl:2018wfy,Gaunt:2018eix,Rinaldi:2018bsf,Rinaldi:2018slz,
Cabouat:2019gtm,Diehl:2019rdh,Diehl:2018kgr,Vladimirov:2016qkd,
Calucci:2008jw,Calucci:2009sv,Calucci:2009ea,Treleani:2012zi,Snigirev:2016uaq,dEnterria:2016ids}, vast phenomenological applications~\cite{Kom:2011bd,Lansberg:2014swa,Lansberg:2015lva,Borschensky:2016nkv,Shao:2016wor,
Lansberg:2016rcx,Lansberg:2016muq,Lansberg:2017chq,Blok:2010ge,Gaunt:2010pi,Maina:2010vh,Berger:2011ep,Kom:2011nu,Luszczak:2011zp,Kasemets:2012pr,Golec-Biernat:2014nsa,vanHameren:2014ava,Echevarria:2015ufa,Blok:2016lmd,Blok:2015afa,Tao:2015nra,Blok:2015rka,Maciula:2015vza,
Luszczak:2014mta,Ceccopieri:2017oqe,Cao:2017bcb,Kumar:2019twx,Cotogno:2018mfv,
Maina:2009sj,Maciula:2017meb,Maciula:2017wpe,Shao:2019qob}, and impressive experimental measurements~\cite{Aaij:2011yc,Abazov:2014qba,Khachatryan:2014iia,Aaboud:2016fzt,
Aaij:2016bqq,Abazov:2015fbl,Khachatryan:2016ydm,Aad:2014rua,Aaboud:2019wfr,Aad:2014kba,
Aaij:2012dz,Aaij:2015wpa,Aad:2013bjm,Chatrchyan:2013xxa,Aad:2016ett,Sirunyan:2017hlu,Sirunyan:2019zox} of DPS, and even triple parton scattering in the last decade, concentrating on proton-proton ($pp$) collisions. Moreover, following the pioneering work~\cite{Strikman:2001gz} by Strikman and Treleani, it is suggested that DPS cross sections in proton-nucleus ($pA$) and nucleus-nucleus ($AB$ or $AA$) collisions will be largely enhanced thanks to collision geometry. The DPS $pA$ cross sections scale by 3 times the number of nucleons $A$ in a nucleus~\cite{Frankfurt:2004kn,Cattaruzza:2004qb,Blok:2012jr,dEnterria:2012jam}, while the single parton scattering (SPS) cross sections in $pA$ only scale by the nuclear mass number $A$ (modulo other nuclear matter effects). The geometrical enhancement is more pronounced in nucleus-nucleus collisions. The DPS cross sections in $AA$ collisions scale as $A^{3.3}/5$~\cite{dEnterria:2013mrp,dEnterria:2014lwk}, while those for SPS are scaling as $A^2$. However, one should bear in mind that such quantitative estimates are based on the assumption that the nuclear matter effects (thermal or nonthermal) are independent of the collision geometry, which is, in fact, not always justified. One counterexample is the nuclear modification of the initial parton flux encoded in the nuclear parton distribution functions (nPDFs). Although the additional geometric effect from the nuclear modifications for DPS is insignificant anyway if the sizes of such modifications are small, the nuclear parton densities can deviate significantly from their free-nucleon counterparts at a scale of a few GeV (see, e.g., Refs.~\cite{Kusina:2017gkz,Lansberg:2016deg,Eskola:2019bgf}). Therefore, the existing DPS formula in heavy-ion collisions should be revised in order to incorporate the extra geometric/spatial effect from nPDFs.~\footnote{To the best of our knowledge, none of the existing DPS phenomenological applications in heavy-ion collisions~\cite{Strikman:2010bg,Blok:2012jr,dEnterria:2012jam,dEnterria:2013mrp,dEnterria:2014lwk,Cazaroto:2016nmu,dEnterria:2017yhd,
Blok:2017alw,Alvioli:2018jls,Alvioli:2019kcy,Helenius:2019uge,Blok:2019fgg,Fedkevych:2019ofc} considers such an effect.}

In addition, the understanding of the spatial (or impact-parameter) dependence of nPDFs is also essential to interpret the nuclear observables measured in different centrality classes, where one usually has to use the (optical or Monte Carlo) Glauber model to link the impact parameter and the centrality. Given the modest amount of available nuclear hard-reaction-process data available in the global fits, almost all considered nPDFs~\cite{Walt:2019slu,AbdulKhalek:2019mzd,Eskola:2016oht,Kovarik:2015cma,Khanpour:2016pph,deFlorian:2011fp,
Eskola:2009uj,Eskola:2008ca,Hirai:2007sx,Hirai:2004wq,deFlorian:2003qf,Hirai:2001np,Eskola:1998df} nowadays are only spatially averaged. They can only be directly used in minimum-bias nuclear collisions. The only one exception is that the authors of Ref.~\cite{Helenius:2012wd} determined the two impact-parameter dependent nPDFs~\footnote{There are also a few attempts to obtain the spatial form of the nPDFs based on phenomenological models (see, e.g., Refs.~\cite{Frankfurt:2011cs,Frankfurt:2016qca,Kumano:1989eh}).} from the $A$ dependencies of the two spatially averaged nPDFs. New means of extracting the spatial dependencies of the nPDFs from experimental data are therefore desired not only because the spatially dependent nPDFs in Ref.~\cite{Helenius:2012wd} are more-or-less obsolete, but also due to the fact that a second independent validation is always valuable.

The primary goal of the paper is to derive the generic expression for the DPS cross section in heavy-ion collisions by  accommodating both the nuclear collision geometry and the spatially dependent nuclear modification effect. Additionally, we also suggest that the measurements of the DPS processes in minimum-bias collisions are able to constrain the impact-parameter dependent nPDFs. The remainder of the context is organized as follows. After introducing the nucleon density and the thickness function in Sec.~\ref{sec:rhoA}, we derive a new generic formula for the DPS cross section in nucleus-nuclues collisions in Sec.~\ref{sec:DPSAB}. We explore the possibility of using the DPS cross sections in $pA$ to determine the spatial dependence of the nPDFs in Sec.~\ref{sec:nPDF}. A short summary is presented in Sec.~\ref{sec:summary}. Finally, Appendix~\ref{sec:tpFpp} discusses the transverse parton profile and the overlap function, and Appendix~\ref{sec:pndiff} considers the case when the transverse position dependencies of protons and neutrons are different in a nucleus.

\medskip


\section{The nucleon density and the thickness function\label{sec:rhoA}}

The nucleon number density in a nucleus is usually parameterized by a Woods-Saxon nucleon density function
\begin{eqnarray}
\rho_A(\overrightarrow{r})&=&\rho_{0,A}\frac{1+w_A(r/R_A)^2}{1+\exp{\left(\frac{r-R_A}{a_A}\right)}},\label{eq:rhoAWS}
\end{eqnarray}
where $r\equiv |\overrightarrow{r}|$, $\rho_{0,A}$ corresponds to the nucleon density in the center of the nucleus, $R_A$ is the radius of the nucleus $A$, $a_A$ is the skin thickness, and $w_A$ characterizes deviations from a spherical shape. The concrete values of these parameters can be found in, e.g., Ref.~\cite{DeJager:1987qc}. Such a function should work well for nuclei with $A\geq 4$. An alternative simpler density function is the so-called hard-sphere model, i.e.,
\begin{eqnarray}
\rho_A(\overrightarrow{r})=\rho_{0,A}\theta(R_A-r),\label{eq:rhoAhardsphere}
\end{eqnarray}
where $\theta(x)$ is the Heaviside function. The normalization for $\rho_A$ is
\begin{eqnarray}
\int{d^3\overrightarrow{r}\rho_A(\overrightarrow{r})}=A.
\end{eqnarray}
For convenience, we also define the nucleon probability density $\hat{\rho}_A(\overrightarrow{r})\equiv \frac{\rho_A(\overrightarrow{r})}{A}$, which is normalized to unity.
In the case of a single nucleon, we can write
\begin{eqnarray}
\rho_{N}(\overrightarrow{r})=\delta^3(\overrightarrow{r}),
\end{eqnarray}
where $\delta^n()$ is an $n$-dimensional Dirac delta function. For other $A<4$ nuclei, one should utilize other nucleon density profiles. For instance, a profile for deuterium was suggested in Ref.~\cite{Helenius:2012wd}.

\begin{figure*}[hbt!]
\centering
\includegraphics[width=1.0\textwidth,draft=false]{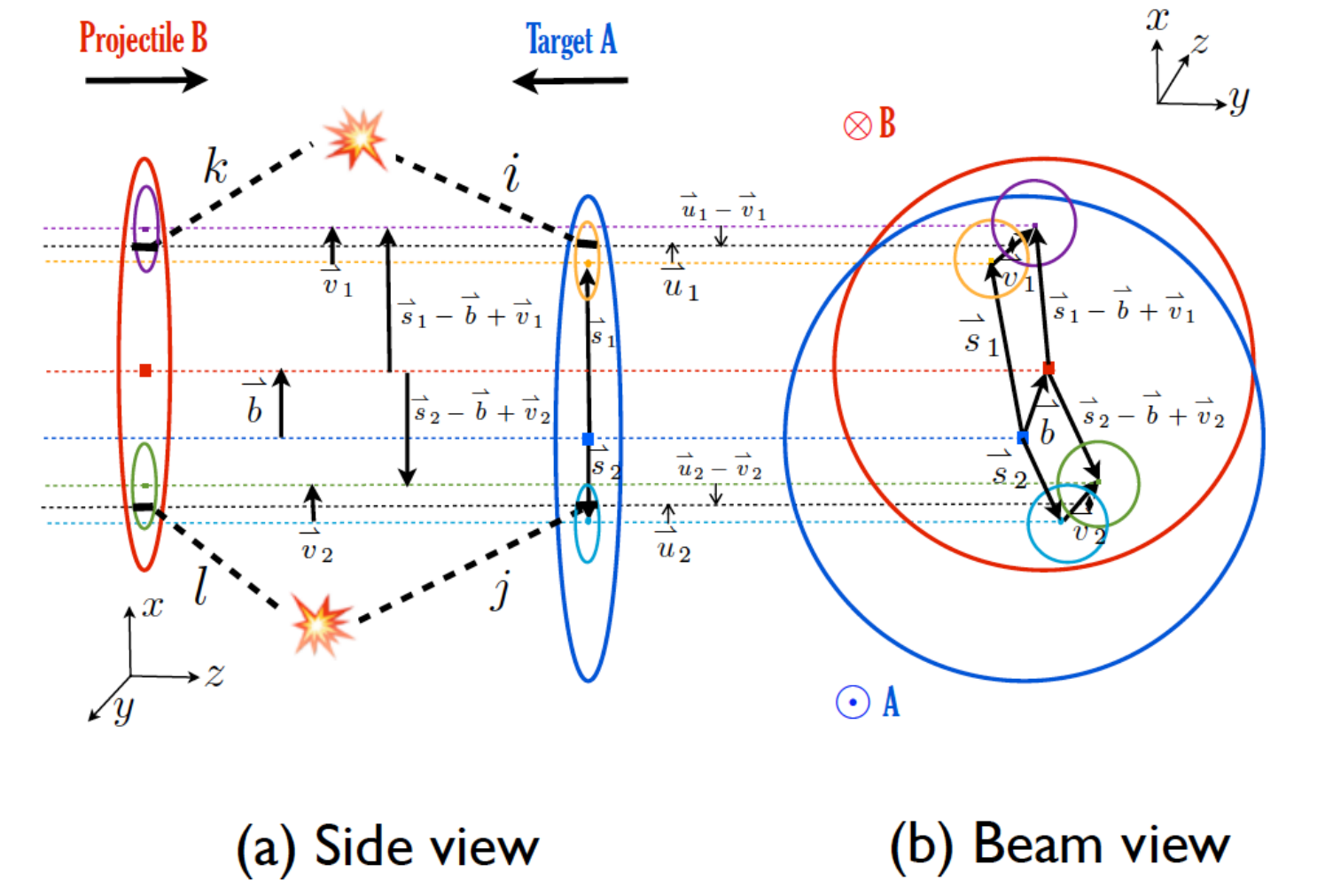}
\caption{\label{fig:geometry} Schematic representations of the geometry for DPS in nucleus-nucleus collisions from (a) side and (b) beam views. The two large ellipses (circles) represent the two colliding nuclei, while the four small ellipses (circles) are the nucleons.}
\end{figure*}

Let us consider a heavy-ion collision of $A$ (the target) and $B$ (the projectile), which is schematically shown in Fig.~\ref{fig:geometry}. Their transverse displacement is a two-dimensional vector $\harpoon b$. One considers two flux tubes located at the transverse displacement $\harpoon s$ and $\harpoon{s}-\harpoon{b}$ with respect to the centers of the target $A$ and the projectile $B$, respectively. In our notation, a three-dimensional vector $\overrightarrow{x}$ can be decomposed into a two-dimensional transverse part $\harpoon{x}$ and  the longitudinal part $x_z$, i.e., $\overrightarrow{x}=(\harpoon{x},x_z)$. The two-dimensional nucleon number density or the thickness function is
\begin{eqnarray}
T_A(\harpoon{b})=\int_{-\infty}^{+\infty}{\rho_A(\harpoon{b},z_A)dz_A},
\end{eqnarray}
where the nucleon probability density per unit transverse area is
\begin{eqnarray}
\hat{T}_A(\harpoon{b})\equiv \frac{T_A(\harpoon{b})}{A} = \int_{-\infty}^{+\infty}{\hat{\rho}_A(\harpoon{b},z_A)dz_A}.
\end{eqnarray}
The integrations of $T_A(\harpoon{b})$ and $\hat{T}_A(\harpoon{b})$ over $\harpoon{b}$ in the whole two-dimensional area result in the nucleon number $A$ and unity. One can define the thickness function $T_{AB}(\harpoon{b})$ and thickness probability function  $\hat{T}_{AB}(\harpoon{b})$ for $AB$ collisions as follows:
\begin{eqnarray}
T_{AB}(\harpoon{b})&=&\int_{-\infty}^{+\infty}{T_A(\harpoon{s})T_B(\harpoon{s}-\harpoon{b})d^2\harpoon{s}},\nonumber\\
\hat{T}_{AB}(\harpoon{b})&=&\int_{-\infty}^{+\infty}{\hat{T}_A(\harpoon{s})\hat{T}_B(\harpoon{s}-\harpoon{b})d^2\harpoon{s}},\label{eq:TABdef}
\end{eqnarray}
which are normalized to $AB$ and unity after integrating over the transverse plane $\harpoon{b}$. One can interpret the thickness probability function $\hat{T}_{AB}(\harpoon{b})$ as the effective overlap area for which a nucleon in $A$ can meet with a nucleon in $B$, which is a pure geometrical factor. 

\medskip

\section{DPS in nucleus-nucleus collisions\label{sec:DPSAB}}

The DPS cross section for a generic reaction $AB\to f_1f_2$ is
\begin{eqnarray}
&&\sigma_{AB\rightarrow f_1 f_2}^{\rm DPS}=\frac{1}{1+\delta_{f_1f_2}}\sum_{i,j,k,l}{\int{dx_1dx_2dx_1^{\prime}dx_2^{\prime}}}\nonumber\\
&&\Gamma^{ij}_{A}(x_1,x_2,\harpoon{s}_1,\harpoon{s}_2,\harpoon{u}_1,\harpoon{u}_2)\hat{\sigma}^{f_1}_{ik}(x_1,x_1^{\prime})\hat{\sigma}^{f_2}_{jl}(x_2,x_2^{\prime})\times\nonumber\\
&&\Gamma_{B}^{kl}(x_1^{\prime},x_2^{\prime},\harpoon{s}_1-\harpoon{b}+\harpoon{v}_1,\harpoon{s}_2-\harpoon{b}+\harpoon{v}_2,\harpoon{u}_1-\harpoon{v}_1,\harpoon{u}_2-\harpoon{v}_2)\nonumber\\
&& d^2\harpoon{u}_1d^2\harpoon{u}_2d^2\harpoon{v}_1d^2\harpoon{v}_2d^2\harpoon{s}_1d^2\harpoon{s}_2d^2\harpoon{b},\label{eq:masterequation}
\end{eqnarray}
where the relevant geometry is shown in Fig.~\ref{fig:geometry}. $\hat{\sigma}^{f_1}_{ik}$ and $\hat{\sigma}^{f_2}_{jl}$ are the two partonic cross sections for $ik\to f_1$ and $jl\to f_2$ with the initial partons $i,j,k,l$ being either (anti)quarks or gluons. $\delta_{f_1f_2}$ is the Kronecker delta function to take into account the symmetry of the final states $f_1$ and $f_2$ in the reaction. There are two contributions for the generalized double parton distribution (GDPD) of the nucleus $A$.~\footnote{The parton distributions also depend on the factorization scale $\mu_F$, which we have neglected in our notations since $\mu_F$ will not affect our discussions.} They are
\begin{eqnarray}
&&\Gamma_A^{ij}(x_1,x_2,\harpoon{s}_1,\harpoon{s}_2,\harpoon{u}_1,\harpoon{u}_2)=\nonumber\\
&&\delta^2(\harpoon{s}_1-\harpoon{s}_2)T_A(\harpoon{s}_1)\bar{\Gamma}_{N/A}^{ij}(x_1,x_2,\harpoon{s}_1,\harpoon{u}_1,\harpoon{u}_2)\nonumber\\
&&+\frac{A-1}{2A}T_A(\harpoon{s}_1)T_A(\harpoon{s}_2)\left[\bar{\Gamma}^i_{N/A}(x_1,\harpoon{s}_1,\harpoon{u}_1)\bar{\Gamma}^j_{N/A}(x_2,\harpoon{s}_2,\harpoon{u}_2)\right.\nonumber\\
&&\left.+ \bar{\Gamma}^i_{N/A}(x_1,\harpoon{s}_2,\harpoon{u}_1)\bar{\Gamma}^j_{N/A}(x_2,\harpoon{s}_1,\harpoon{u}_2)\right]\label{eq:gammaAij}
\end{eqnarray}
where $\bar{\Gamma}^{ij}_{N/A}$ and $\bar{\Gamma}^i_{N/A}$ are the isospin-averaged GDPD and the isospin-averaged generalized single parton distribution (GSPD) for a nucleon in $A$. The isospin average is written as
\begin{eqnarray}
\bar{\Gamma}^{ij}_{N/A}&=&\overline{\sum_{N^A}}{\Gamma^{ij}_{N^A}}=\frac{1}{A}\sum_{N^A}{\Gamma^{ij}_{N^A}}=\frac{Z}{A}\Gamma^{ij}_{p^A}+\frac{A-Z}{A}\Gamma^{ij}_{n^A},\nonumber\\
\bar{\Gamma}^i_{N/A}&=&\overline{\sum_{N^A}}{\Gamma^{i}_{N^A}}=\frac{1}{A}\sum_{N^A}{\Gamma^{i}_{N^A}}=\frac{Z}{A}\Gamma^{i}_{p^A}+\frac{A-Z}{A}\Gamma^{i}_{n^A},\nonumber\\
\end{eqnarray}
for a nucleus $A$ with $Z$ protons and $A-Z$ neutrons. We call $\Gamma^{ij}_{N^A}$ and $\Gamma^{i}_{N^A}$ the generalized double parton distribution in a nucleon and the generalized single parton distribution in a nucleon, respectively. We have used the bounded nucleon (bounded proton, bounded neutron) in nucleus $A$ as $N^A$ ($p^A$, $n^A$), while a free-nucleon, a free-proton, and a free-neutron are denoted as $N, p$, and $n$, respectively.
The first term on the right-hand side of Eq. (\ref{eq:gammaAij}) represents the two partons $i$ and $j$, which belong to the same nucleon in nuclues $A$ with the impact parameter of the nucleon $\harpoon{s}_1$. The second term means that the two partons are from two distinct nucleons, where the prefactor $\frac{A-1}{A}$ takes into account the difference between the number of nucleon pairs and the number of different nucleon pairs. Such a factor is essential to guarantee the correct normalization of $\Gamma_A^{ij}$, which was first noticed in Refs.~\cite{Frankfurt:2004kn,Blok:2012jr}. If $A$ is a nucleon ($A=1$), the second term is zero because of the prefactor. Finally, we also have the decomposition for the GDPD $\Gamma_{B}^{kl}$ of nucleus $B$ akin to Eq. (\ref{eq:gammaAij}) for nucleus $A$.

We can use the factorized ansatz for the remaining nucleon GDPD $\Gamma_{N^A}^{ij}$ and the nucleon GSPDs $\Gamma^i_{N^A}$ and $\Gamma^j_{N^A}$ as follows:
\begin{eqnarray}
&&\Gamma_{N^A}^{ij}(x_1,x_2,\harpoon{s}_1,\harpoon{u}_1,\harpoon{u}_2)\nonumber\\
&&=t_{N^A}(\harpoon{u}_1)t_{N^A}(\harpoon{u}_2)g_{N^A}^i(x_1,\harpoon{s}_1)g_{N^A}^j(x_2,\harpoon{s}_1),\nonumber\\
&&\Gamma^i_{N^A}(x_1,\harpoon{s}_1,\harpoon{u}_1)=t_{N^A}(\harpoon{u}_1)g_{N^A}^i(x_1,\harpoon{s}_1),\nonumber\\
&&\Gamma^j_{N^A}(x_2,\harpoon{s}_2,\harpoon{u}_2)=t_{N^A}(\harpoon{u}_2)g_{N^A}^j(x_2,\harpoon{s}_2),\label{eq:gammafact}
\end{eqnarray}
where we have used $t_{N^A}(\harpoon{u})$ as the transverse parton profile in the bounded nucleon $N^A$ and $g_{N^A}^i(x,\harpoon{s})$ the impact-parameter $\harpoon{s}$ dependent nPDF for the parton $i$. A similar factorized ansatz is widely used in DPS processes in (free) nucleon-nucleon collisions. It assumes the vanishing parton-parton correlations in DPS and yields the well-known ``pocket formula" for the cross section in nucleon-nucleon $N_1N_2$ collisions
\begin{eqnarray}
\sigma^{\rm DPS}_{N_1N_2\to f_1f_2}&=&\frac{1}{1+\delta_{f_1f_2}}\frac{\sigma_{N_1N_2\to f_1}\sigma_{N_1N_2\to f_2}}{\sigma_{{\rm eff},N_1N_2}}
\end{eqnarray}
with the effective cross section as
\begin{eqnarray}
\sigma_{{\rm eff},N_1N_2}&=&\left[\int{F_{N_1N_2}(\harpoon{v})d^2\harpoon{v}}\right]^{-1}
\end{eqnarray}
and the overlap function
\begin{eqnarray}
F_{N_1N_2}(\harpoon{v})&=&\int{t_{N_1}(\harpoon{u})t_{N_2}(\harpoon{u}-\harpoon{v})d^2\harpoon{u}}.\label{eq:overlapF}
\end{eqnarray}
Usually, one assumes that the transverse parton profile $t_N(\harpoon{u})$ is independent of the type of the free-nucleon $N$, which is either a proton or a neutron. Then, we are left with one single effective cross section parameter $\sigma_{{\rm eff}, N_1N_2}=\sigma_{{\rm eff}, pp}, \forall\ N_i \in \{p,n\}$. In addition, it is reasonable to assume that the transverse parton profile is not affected by the surrounding nucleons in a nucleus, i.e., $t_{N^A}(\harpoon{u})=t_{N}(\harpoon{u})$. In the following, we will use such two simplifications and retain only one $t_p$ and one $F_{pp}$ as the unique transverse parton profile and the overlap function. Further discussions about these two functions can be found in Appendix \ref{sec:tpFpp}.

Motivated by the shadowing at small $x$ and the Gribov-Glauber modeling~\cite{Frankfurt:2011cs,Armesto:2010kr,Tywoniuk:2007xy} of the nPDFs, we can assume the nuclear matter effects encoded in nPDFs are only depending on the thickness function $T_A$. We can introduce the general expression as
\begin{eqnarray}
\frac{g^i_{N^A}(x,\harpoon{s})}{g^i_N(x)}-1=\left(\frac{g^i_{N^A}(x)}{g^i_N(x)}-1\right)G\left(\frac{T_A(\harpoon{s})}{T_A(\harpoon{0})}\right),\label{eq:bnPDF}
\end{eqnarray}
where $g^i_N(x)$ is the free-nucleon $N$ PDF for parton $i$ and $g^i_{N^A}(x)$ is the spatially averaged nucleon $N$ PDF for parton $i$ in $A$. $G()$ can be an arbitrary function~\footnote{A reasonable constraint one can impose is $\lim_{|\harpoon{s}|\to +\infty}{G\left(\frac{T_A(\harpoon{s})}{T_A(\harpoon{0})}\right)}=0$. This can be understood because, at sufficiently large distance, the nucleons would behave like free particles. We, however, will not use such a constraint in the following discussions.} with the normalization condition
\begin{eqnarray}
\int{T_A(\harpoon{s})G\left(\frac{T_A(\harpoon{s})}{T_A(\harpoon{0})}\right)d^2\harpoon{s}}&=&A.
\end{eqnarray}
The simple form $G\left(\frac{T_A(\harpoon{s})}{T_A(\harpoon{0})}\right)=\frac{AT_A(\harpoon{s})}{T_{AA}(\harpoon{0})}$ is the most frequently used one in the literature~\cite{Eskola:1991ec,Emelyanov:1999pkc,Klein:2003dj,Vogt:2004dh,Vogt:2004hf,Ferreiro:2008wc,
Vogt:2010aa} and also in the {\sc\small hijing} event generator~\cite{Gyulassy:1994ew}. However, such a simple form conflicts with the $A$ dependence of the nPDF global fit~\cite{Helenius:2012wd}. A study based on the $A$ dependencies of the nPDFs reveals that a polynomial function $G()$ with terms up to $\left(\frac{T_A(\harpoon{s})}{T_A(\harpoon{0})}\right)^4$ can reproduce the nPDF $A$ dependence over the entire $x$ range. In the following, for simplicity, we will use the abbreviations $G_{A,1}(\harpoon{s})\equiv G\left(\frac{T_A(\harpoon{s})}{T_A(\harpoon{0})}\right)$, $G_{A,2}(\harpoon{s})\equiv 1-G_{A,1}(\harpoon{s})$, $g^i_{N^A,1}(x)\equiv g^i_{N^A}(x)$, and $g^i_{N^A,2}(x)\equiv g^i_{N}(x)$. Therefore, Eq.~(\ref{eq:bnPDF}) can be reformulated as
\begin{eqnarray}
g^i_{N^A}(x,\harpoon{s})&=&\sum_{m=1}^{2}{g^i_{N^A,m}(x)G_{A,m}(\harpoon{s})}.\label{eq:bnPDF2}
\end{eqnarray}

After applying the ansatz Eq.~(\ref{eq:gammafact}) and the relation Eq.~(\ref{eq:bnPDF2}) into Eq.~(\ref{eq:masterequation}), we arrive at the main result of the paper
\begin{eqnarray}
&&\sigma^{\rm DPS}_{AB\to f_1f_2}=\frac{1}{1+\delta_{f_1f_2}}\sum_{N_1^A,N_2^A,N_1^B,N_2^B}{\sum_{m_1,m_2,m_3,m_4=1}^{2}}\nonumber\\
&&\times \left(\sigma^{m_1m_3}_{N_1^AN_1^B\to f_1}\sigma^{m_2m_4}_{N_2^AN_2^B \to f_2}\right)\nonumber\\
&&\times \left[\delta_{N_1^AN_2^A}\delta_{N_1^BN_2^B}\frac{\hat{T}_{A,m_1m_2}\hat{T}_{B,m_3m_4}}{\sigma_{{\rm eff},pp}}\right.\nonumber\\
&&+\delta_{N_1^BN_2^B}\frac{A-1}{A}\hat{T}^{(2)}_{nA,m_1m_2}\hat{T}_{B,m_3m_4}\nonumber\\
&&+\delta_{N_1^AN_2^A}\frac{B-1}{B}\hat{T}_{A,m_1m_2}\hat{T}^{(2)}_{nB,m_3m_4}\nonumber\\
&&\left.+\frac{(A-1)(B-1)}{AB}\hat{T}^{(2)}_{nAB,m_1m_2m_3m_4}\right],\label{eq:mainAB}
\end{eqnarray}
where we have used
\begin{eqnarray}
&&\sigma^{m_1m_3}_{N_1^AN_1^B\to f_1}=\nonumber\\
&&\sum_{i,k}{\int{dx_1dx_1^\prime g^i_{N_1^A,m_1}(x_1)g^k_{N_1^B,m_3}(x_1^\prime)\hat{\sigma}_{ik}^{f_1}(x_1,x_1^\prime)}},\nonumber\\
&&\sigma^{m_2m_4}_{N_2^AN_2^B\to f_2}=\nonumber\\
&&\sum_{j,l}{\int{dx_2dx_2^\prime g^j_{N_2^A,m_2}(x_2)g^l_{N_2^B,m_4}(x_2^\prime)\hat{\sigma}_{jl}^{f_2}(x_2,x_2^\prime)}}.
\end{eqnarray}
In addition, the symbols $\hat{T}$ are defined as
\begin{eqnarray}
&&\hat{T}_{A,m_1m_2}=\int{\hat{T}_A(\harpoon{b})G_{A,m_1}(\harpoon{b})G_{A,m_2}(\harpoon{b})d^2\harpoon{b}},\nonumber\\
&&\hat{T}_{nA,m_1m_2}^{(2)}=\int{\left(\hat{T}_{nA,m_1}(\harpoon{b})\hat{T}_{nA,m_2}(\harpoon{b})\right)d^2\harpoon{b}},\nonumber\\
&&\hat{T}^{(2)}_{nAB,m_1m_2m_3m_4}=\frac{1}{2}\int{\left[\hat{T}_{nAB,m_1m_3}(\harpoon{b})\hat{T}_{nAB,m_2m_4}(\harpoon{b})\right.}\nonumber\\
&&\left.+\hat{T}_{nAB,m_1m_4}(\harpoon{b})\hat{T}_{nAB,m_2m_3}(\harpoon{b})\right]d^2\harpoon{b},\nonumber\\
&&\hat{T}_{nA,m}(\harpoon{b})=\int{F_{pp}(\harpoon{v})\hat{T}_A(\harpoon{v}-\harpoon{b})G_{A,m}(\harpoon{v}-\harpoon{b})d^2\harpoon{v}},\nonumber\\
&&\hat{T}_{nAB,m_1m_2}(\harpoon{b})=\int{F_{pp}(\harpoon{v})\hat{T}_{AB,m_1m_2}(\harpoon{b}-\harpoon{v})d^2\harpoon{v}},\nonumber\\
&&\hat{T}_{AB,m_1m_2}(\harpoon{b})=\int{\left[\hat{T}_A(\harpoon{s})G_{A,m_1}(\harpoon{s})\right.}\nonumber\\
&&\left.\times\hat{T}_B(\harpoon{s}-\harpoon{b})G_{B,m_2}(\harpoon{s}-\harpoon{b})\right]d^2\harpoon{s}.
\end{eqnarray}
Because of the normalization relations, we have
\begin{eqnarray}
\hat{T}_{A,12}&=&\hat{T}_{A,21}=-\hat{T}_{A,22}=1-\hat{T}_{A,11}.
\end{eqnarray}
The interpretation of the four terms in the brackets of Eq.~(\ref{eq:mainAB}) is straightforward. They represent three different DPS contributions from nucleus-nucleus interactions. The first term is from the two pairs of the colliding partons belonging to the same pair of incident nucleons. The second and the third terms originate from the two partons from a nucleon in a nucleus interaction with the two partons from two different nucleons in another nucleus. The last term is the contribution of the two pairs of partons belonging to two different nucleons from both nuclei.

A few special situations are worth being explored. When we take the identity of the impact-parameter dependent nPDF $g^i_{N^A}(x,\harpoon{s})$ and the spatially averaged nPDF $g^i_{N^A}(x)$ via $G_{A,1}(\harpoon{s})=1$ and $G_{A,2}(\harpoon{s})=0$, we can recover the well-known DPS formula in $AB$ collisions [see, e.g., Eqs.(1) and (2) in Ref.~\cite{Helenius:2019uge}], which, however, does not take into account the spatially dependent initial nuclear modifications. Moreover, if we set $g^i_{N^A}(x)=g^i_{N}(x)$ (i.e., zero nuclear modification), we have $g_{N^A,1}^i(x)=g_{N^A,2}^i(x)=g^i_{N}(x)$. The final expression is independent of $G()$, as it must be. Finally, if we take nucleus $B$ as a proton, which amounts to setting $B=1,G_{B,1}(\harpoon{s})=1$, and $G_{B,2}(\harpoon{s})=0$, Eq.~(\ref{eq:mainAB}) is reduced to
\begin{eqnarray}
&&\sigma^{\rm DPS}_{Ap\to f_1f_2}=\frac{1}{1+\delta_{f_1f_2}}\sum_{N_1^A,N_2^A}{\sum_{m_1,m_2=1}^{2}}\nonumber\\
&&\times \left(\sigma^{m_1 1}_{N_1^A p\to f_1}\sigma^{m_2 1}_{N_2^A p \to f_2}\right)\nonumber\\
&&\times \left[\delta_{N_1^AN_2^A}\frac{\hat{T}_{A,m_1m_2}}{\sigma_{{\rm eff},pp}}+\frac{A-1}{A}\hat{T}^{(2)}_{nA,m_1m_2}\right].\label{eq:mainAp}
\end{eqnarray}
This gives rise to a DPS formula in $pA$ (or $Ap$) collisions.

If $A\gg 1$, we can impose a good approximation $F_{pp}(\harpoon{v})\approx \delta^2(\harpoon{v})$. This can be understood because a nucleon in a heavy nucleus looks like a point in space. Then,
\begin{eqnarray}
\hat{T}_{nA,m}(\harpoon{b})&\approx& \hat{T}_A(\harpoon{b})G_{A,m}(\harpoon{b}),\nonumber\\
\hat{T}_{nAB,m_1m_2}(\harpoon{b})&\approx& \hat{T}_{AB,m_1m_2}(\harpoon{b}).\label{eq:Fvapprox}
\end{eqnarray}
With such a simplification, the transverse parton profile $t_p$ will only enter into $\sigma_{{\rm eff},pp}$ in Eqs.~(\ref{eq:mainAB}) and (\ref{eq:mainAp}). The goodness for the above approximation will be validated in Appendix \ref{sec:tpFpp} with a few concrete modelings of $F_{pp}(\harpoon{v})$.

Finally, it would be useful to consider a few exceptional cases in which our assumptions do not hold. The first case is when our factorization ansatz Eq.~(\ref{eq:bnPDF}) is violated by, for instance, the existence of strong correlations~\cite{Alvioli:2014eda}. The concrete formulas (\ref{eq:mainAB}) and (\ref{eq:mainAp}) for DPS cross sections should be revised depending on the new ansatz. Our general idea of using DPS cross sections in minimum-bias nuclear collisions as a sensitive probe of the transverse position dependence of the nuclear modification of parton densities is, however, still valid. For simplicity, we have not discriminated the possible different transverse position dependencies of protons and neutrons in nuclei, also known as the neutron skin effect (see, e.g., in Refs.~\cite{Abrahamyan:2012gp,Tarbert:2013jze,Paukkunen:2015bwa}). Since the first sums in Eqs.~(\ref{eq:mainAB}) and (\ref{eq:mainAp}) run over all possible (bounded) nucleons, it is easy to incorporate such an effect, which can be found in Appendix \ref{sec:pndiff}.

\medskip

\section{Impact-parameter dependent nuclear PDF from DPS\label{sec:nPDF}}

From Eqs.~(\ref{eq:mainAB}) and (\ref{eq:mainAp}), we know that the DPS cross sections in nuclear collisions depend on the function $G()$ characterizing the impact-parameter dependence of nPDFs, as introduced in Eq.~(\ref{eq:bnPDF}). In turn, we can view DPS as a probe to determine the spatial dependence $G()$. The task of DPS cross section extraction from experimental data is, however, far from nontrivial due to the presence of the contamination from the SPS contribution. An ideal case is to look for a process in which the SPS contribution is suppressed. A few such examples are same-sign open charm~\cite{Aaij:2012dz}, $J/\psi+$charm~\cite{Aaij:2012dz}, $\Upsilon+$charm~\cite{Aaij:2015wpa},~\footnote{A recent calculation based on $k_T$ factorization~\cite{Karpishkov:2019vyt} shows that SPS is very big in $\Upsilon+$charm production.} and $J/\psi+\Upsilon$~\cite{Shao:2016wor,Abazov:2015fbl} production. In order to avoid the complications from the final-state nuclear effects, we only take the $pA$ collisions as an example here. The above mentioned processes are dominated by gluon-gluon initial state at the LHC energies, which is blind with the isospin effect. For these gluon-induced processes, Eq.~(\ref{eq:mainAp}) can be further simplified. The nuclear modification factor is expressed as
\begin{eqnarray}
&&R^{\rm DPS}_{pA \to f_1f_2}\equiv \frac{\sigma_{pA \to f_1f_2}^{\rm DPS}}{A\sigma_{pp\to f_1f_2}^{\rm DPS}}\nonumber\\
&=&\sum_{i,j=1}^2{\left(\hat{T}_{A,ij}+(A-1)\sigma_{{\rm eff},pp}\hat{T}^{(2)}_{nA,ij}\right)\left(R_{pA}^{f_1}\right)^{2-i}\left(R_{pA}^{f_2}\right)^{2-j}}\label{eq:RpAgg}
\end{eqnarray}
with $R_{pA}^{f}\equiv \frac{\sigma_{pA\to f}}{A\sigma_{pp\to f}}$ in the minimum-bias collisions.

One reasonable approximation we can take is that the nucleon number density follows the hard-sphere form of Eq.~(\ref{eq:rhoAhardsphere}). Then, the thickness function is $T_A(\harpoon{b})=\frac{3A}{2\pi R_A^2}\sqrt{1-|\harpoon{b}|^2/R_A^2}\theta(R_A-|\harpoon{b}|)$. For illustration purposes only, we consider $G(x)$ as a monomial in the argument $x$ only here, although its practical form can be sufficiently complicated. Therefore, the analytical expression of $G_{A,1}$ is $G_{A,1}(\harpoon{b})=\frac{a+3}{3}\left(\frac{T_A(\harpoon{b})}{T_A(\harpoon{0})}\right)^a$. Then, we can derive
\begin{eqnarray}
&&\hat{T}_{A,11}\approx\frac{3^{1-2a}\left(a+3\right)^{2a}}{2a+3},\nonumber\\
&&\hat{T}_{nA,11}^{(2)}\approx\frac{9^{1-a}(a+3)^{2a}}{4(a+2)\pi R_A^2},\nonumber\\
&&\hat{T}_{nA,12}^{(2)}=\hat{T}_{nA,21}^{(2)}\approx\frac{3^{2-a}(a+3)^{a}}{2\left(a+4\right)\pi R_A^2}-\frac{9^{1-a}(a+3)^{2a}}{4\left(a+2\right)\pi R_A^2},\nonumber\\
&&\hat{T}_{nA,22}^{(2)}\approx\frac{9}{8\pi R_A^2}-\frac{3^{2-a}(a+3)^{a}}{\left(a+4\right)\pi R_A^2}+\frac{9^{1-a}(a+3)^{2a}}{4\left(a+2\right)\pi R_A^2}.
\end{eqnarray}
In such a case, the nuclear modification factor becomes
\begin{eqnarray}
&&R_{pA \to f_1f_2}^{\rm DPS}\approx R_{pA}^{f_1}R_{pA}^{f_2}\left(\frac{3^{1-2a}(a+3)^{2a}}{2a+3}\right.\nonumber\\
&&\left.+\sigma_{{\rm eff},pp}\frac{(A-1)9^{1-a}(a+3)^{2a}}{4(a+2)\pi R_A^2}\right)\nonumber\\
&&+\left(R_{pA}^{f_1}+R_{pA}^{f_2}\right)\left[1-\frac{3^{1-2a}(a+3)^{2a}}{2a+3}\right.\nonumber\\
&&\left.+\sigma_{{\rm eff},pp}\left(A-1\right)\left(\frac{3^{2-a}(a+3)^a}{2(a+4)\pi R_A^2}-\frac{9^{1-a}(a+3)^{2a}}{4(a+2)\pi R_A^2}\right)\right]\nonumber\\
&&+\left[\frac{3^{1-2a}(a+3)^{2a}}{2a+3}-1+\sigma_{{\rm eff},pp}\left(A-1\right)\right.\nonumber\\
&&\left.\times\left(\frac{9}{8\pi R_A^2}+\frac{9^{1-a}(a+3)^{2a}}{4(a+2)\pi R_A^2}-\frac{3^{2-a}(a+3)^a}{(a+4)\pi R_A^2}\right)\right].
\end{eqnarray}
It is easy to check that, when $a=0$ (zero spatial dependence), we are left with the first term proportional to $R_{pA}^{f_1}R_{pA}^{f_2}$. 

\begin{figure}[hbt!]
\centering
\includegraphics[width=1.0\columnwidth,draft=false]{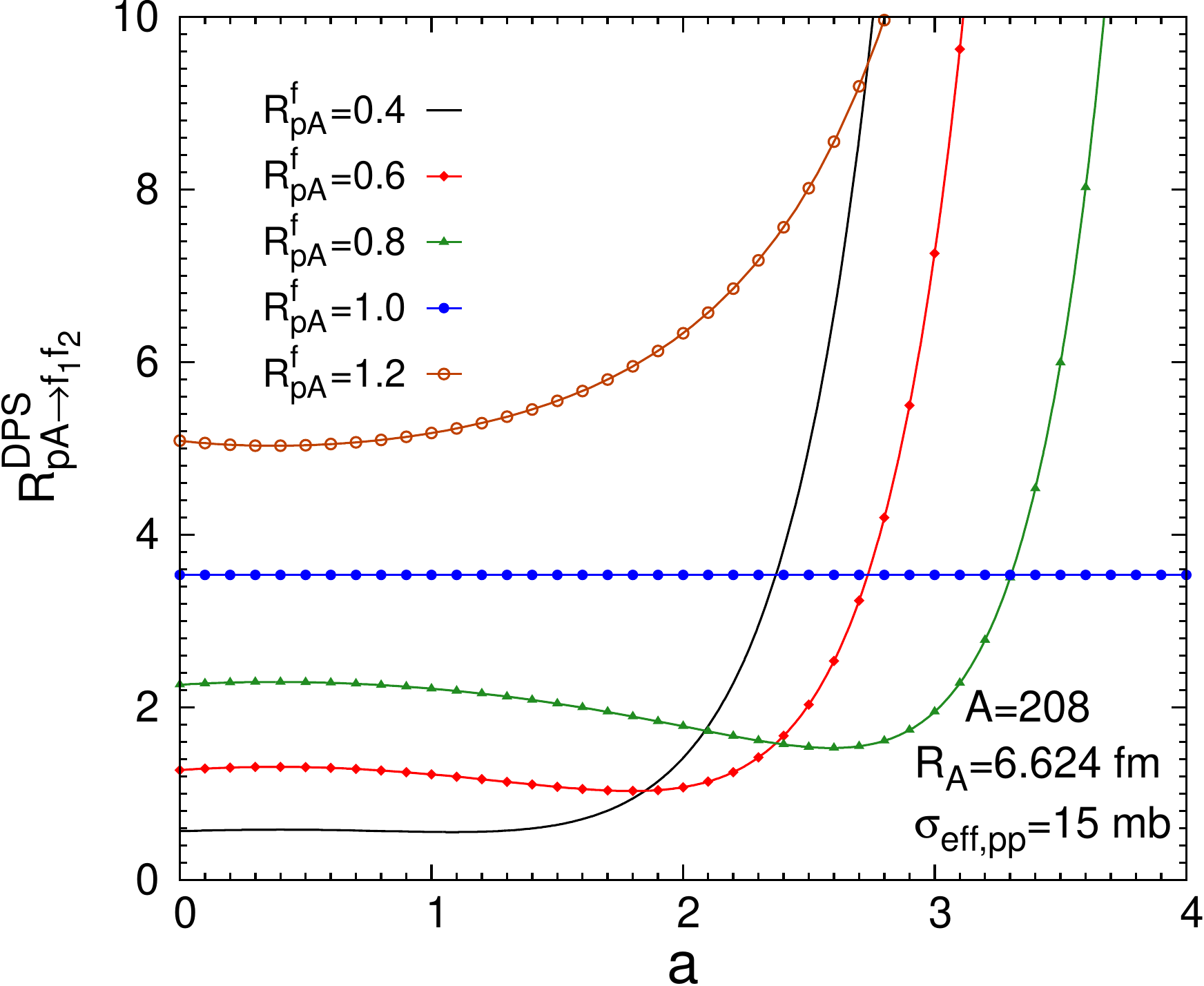}
\caption{\label{fig:RpADPS} $R_{pA\to f_1f_2}^{\rm DPS}$ dependence of $a$, where $a$ is the power of $x$ in $G(x)$ via $G(x)\propto x^a$. Five different exemplified values $R_{pA}^f=R_{pA}^{f_1}=R_{pA}^{f_2}$ are shown.}
\end{figure}

Let us take the lead (Pb) beam with $A=208$, $R_A=6.624$ fm, and $\sigma_{{\rm eff},pp}=15$ mb as a special example. Such a beam is available at the LHC. Different numbers of the power $a$ in $G(x)\propto x^a$ predict quite different values of the nuclear modification factor $R_{pA \to f_1f_2}^{\rm DPS}$, as reported in Fig.~\ref{fig:RpADPS}. The curves corresponding to five different values of $R_{pA}^f=R_{pA}^{f_1}=R_{pA}^{f_2}$ are displayed. $R_{pA\to f_1f_2}^{\rm DPS}$ dramatically increases when $a>1.5, 2.0, 3.0$ and $1.0$ for $R_{pA}^f=0.4,0.6,0.8$, and $1.2$. As anticipated, the curve of $R_{pA}^f=1.0$ (no nuclear modification) is independent of $G(x)$ (or $a$). As realistic examples, $R_{pA}^f$ from the single-$f$ inclusive processes, with $f$ being either the open charm or $J/\psi$ mesons at the LHC proton-lead collisions were precisely measured to be close to $0.6$ in the forward rapidity region (see, e.g., Fig.1 in Ref.~\cite{Kusina:2017gkz}). The $a=0,1,2$, and $3$ predict $R_{pA\to f_1f_2}^{\rm DPS}=1.27, 1.22, 1.07$, and $7.26$ in the same kinematic regime. These numbers can be refined by using the Woods-Saxon density [cf. Eq.~(\ref{eq:rhoAWS})] and with a concrete parton overlap function $F_{pp}(\harpoon{v})$ [cf. Eq.~(\ref{eq:overlapF})]. The numerical differences with respect to what we have shown should be minor though. From this example, we have clearly shown that the nuclear modification factors of $J/\psi$ plus open charm and same-sign charm production in proton-lead collisions will provide precious inputs for determining the impact-parameter dependent nPDFs. Such measurements are independent of the centrality-based measurements, where the latter ones are crucially dependent on Glauber modeling (see, e.g., Refs.~\cite{Vogt:1999jp,Miller:2007ri}) and are subject to large uncertainties, particularly in proton-nucleus collisions. 

\medskip






\section{Summary\label{sec:summary}}

In this paper, for the first time, we have considered both the nuclear collision geometry and the impact-parameter dependent nuclear modification in the nPDFs for DPS processes in heavy-ion collisions. A simple generic equation (\ref{eq:mainAB}) has been derived for evaluating the DPS cross sections in nucleus-nucleus collisions, while its $pA$ counterpart is given in Eq.~(\ref{eq:mainAp}). Both of the above effects are important in scrutinizing the DPS heavy-ion data. The latter is particular relevant when the nuclear modification encoded in the nPDFs is significant (e.g. the open/hidden charm and beauty production~\cite{Kusina:2017gkz} at the LHC). In turn, we can also extract the spatial dependence of the nPDFs by measuring DPS cross sections in minimum-bias nuclear collisions. We take the gluon-induced charm and beauty production processes as an example. $\sigma_{{\rm eff},pp}$ can be determined from their $pp$ data (e.g., Ref.~\cite{Aaij:2012dz}), and $R_{pA}^{f_1}$ and $R_{pA}^{f_2}$ are measured in their single inclusive processes. The measurements of the DPS nuclear modification factor $R^{\rm DPS}_{pA\to f_1f_2}$ can be readily used to pin down the spatial function $G()$ entering into the impact-parameter dependent nPDFs. Such an approach has the virtue of independence of Glauber modeling.

\acknowledgments
I would like to thank Michael Winn for the useful comments on the manuscript.
The work is supported by
the ILP Labex (ANR-11-IDEX-0004-02, ANR-10-LABX-63). 

\appendix

\section{The transverse parton profile and the overlap function\label{sec:tpFpp}}

Several empirical functional forms of the transverse parton profile $t_p(\harpoon{u})$ in a nucleon were suggested in the literature~\cite{Sjostrand:1987su,Frankfurt:2003td,Domdey:2009bg,Gaunt:2014rua}. They are collected in Table~\ref{tab:tnprof}. The ``dipole" profile is equivalent to the ``exponential" profile as long as we take $r_0^{-1}=m_g$. Both of them are proportional to the modified Bessel function $K_1()$. The analytical expressions of the mean three-dimensional radius squared $\langle \overrightarrow{r}^2\rangle$ and the mean two-dimensional radius squared $\langle \harpoon{u}^2\rangle$ can be found in Table~\ref{tab:tnr2avg}. Due to the spatial symmetry, we always have $\langle \overrightarrow{r}^2\rangle=\frac{3}{2}\langle \harpoon{u}^2\rangle$. We have also evaluated the analytic functions of the overlap function $F_{pp}(\harpoon{v})$ and of $\sigma_{{\rm eff},pp}$ for all these profiles in Table~\ref{tab:Fn}.

\begin{table*}[hbt!]
    \centering
    \begin{tabular}{|c|c|c|}
    \hline
      Full name  &  Acronym & Functional form \\\hline
    Hard sphere  &  HS  &   $t_p(\harpoon{u})=\frac{3}{2\pi r_0^2}\sqrt{1-|\harpoon{u}|^2/r_0^2}\theta(r_0-|\harpoon{u}|)$\\
     Gaussian &  G  &   $t_p(\harpoon{u})=\frac{1}{2\pi r_0^2}\exp{\left(-\frac{|\harpoon{u}|^2}{2r_0^2}\right)}$\\
    Double Gaussian & DG & $t_p(\harpoon{u})=\frac{1-\beta}{\pi r_{0,1}^2}\exp{\left(-\frac{|\harpoon{u}|^2}{r_{0,1}^2}\right)}+\frac{\beta}{\pi r_{0,2}^2}\exp{\left(-\frac{|\harpoon{u}|^2}{r_{0,2}^2}\right)}$ \\
     Top hat   &  TH & $t_p(\harpoon{u})=\frac{1}{\pi r_0^2}\theta(r_0-|\harpoon{u}|)$\\
      \multirow{2}{*}{Dipole}    &   \multirow{2}{*}{D}  & $t_p(\harpoon{u})=\int{\frac{d^2\harpoon{\Delta}}{4\pi^2}e^{i \harpoon{\Delta}\cdot \harpoon{u}}\left(|\harpoon{\Delta}|^2/m_g^2+1\right)^{-2}}$\\
     & & $=\int{\frac{|\harpoon{\Delta}|d|\harpoon{\Delta}|}{2\pi}J_0\left(|\harpoon{\Delta}||\harpoon{u}|\right)\left(|\harpoon{\Delta}|^2/m_g^2+1\right)^{-2}}=\frac{m_g^2}{2\pi}\frac{m_g|\harpoon{u}|}{2}K_1(m_g|\harpoon{u}|)$ \\
      Exponential & E & $t_p(\harpoon{u})=\int{\frac{dz}{8\pi r_0^3}\exp{\left(-\frac{\sqrt{|\harpoon{u}|^2+z^2}}{r_0}\right)}}=\frac{1}{2\pi r_0^2}\frac{|\harpoon{u}|}{2r_0}K_1\left(\frac{|\harpoon{u}|}{r_0}\right)$ \\
     \hline
    \end{tabular}
    \caption{A summary of the transverse parton profile in a nucleon.}
    \label{tab:tnprof}
\end{table*}

\begin{table*}[hbt!]
    \centering
    \begin{tabular}{|c|c|c|}
    \hline
      Profile & $\langle \overrightarrow{r}^2 \rangle$ & $\langle \harpoon{u}^2 \rangle$\\\hline
    HS  &   $\frac{3}{5}r_0^2$ & $\frac{2}{5}r_0^2$\\
     G  &   $3r_0^2$ & $2r_0^2$\\
    DG & $\frac{3}{2}\left[(1-\beta)r_{0,1}^2+\beta r_{0,2}^2\right]$ & $(1-\beta)r_{0,1}^2+\beta r_{0,2}^2$ \\
     TH &  $\frac{3}{4}r_0^2$ & $\frac{1}{2}r_0^2$ \\
      D  & $\frac{12}{m_g^2}$ & $\frac{8}{m_g^2}$\\
      E & $12r_0^2$ & $8r_0^2$\\
     \hline
    \end{tabular}
    \caption{The mean three-dimensional radius squared and the mean two-dimensional radius squared.}
    \label{tab:tnr2avg}
\end{table*}

\begin{table*}[hbt!]
    \centering
    \begin{tabular}{|c|c|c|}
    \hline
      Profile & Overlap function & $\sigma_{{\rm eff},pp}$\\\hline
   \multirow{2}{*}{HS}  &   $F_{pp}(\harpoon{v})=\frac{9}{512\pi r_0^6}\left[4r_0\left(8r_0^2+|\harpoon{v}|^2\right)\sqrt{4r_0^2-|\harpoon{v}|^2}\right.$ & \multirow{2}{*}{$\frac{1400\pi}{9\left(179-128\ln{2}\right)}r_0^2$} \\
   & $\left.+|\harpoon{v}|^2\left(16r_0^2-|\harpoon{v}|^2\right)\ln{\frac{2r_0-\sqrt{4r_0^2-|\harpoon{v}|^2}}{2r_0+\sqrt{4r_0^2-|\harpoon{v}|^2}}}\right]\theta(2r_0-|\harpoon{v}|)$ & \\
     G  &   $F_{pp}(\harpoon{v})=\frac{1}{4\pi r_0^2}\exp{\left(-\frac{|\harpoon{v}|^2}{4r_0^2}\right)}$ & $8\pi r_0^2$\\
    \multirow{2}{*}{DG} & $F_{pp}(\harpoon{v})=\frac{\left(1-\beta\right)^2}{2\pi r_{0,1}^2}e^{-\frac{|\harpoon{v}|^2}{2r_{0,1}^2}}+\frac{\beta^2}{2\pi r_{0,2}^2}e^{-\frac{|\harpoon{v}|^2}{2r_{0,2}^2}}$ & $\frac{\pi}{\sum_{i=0}^{4}{\frac{4!}{i!(4-i)!}\frac{(1-\beta)^i\beta^{4-i}}{i r_{0,1}^2+(4-i)r_{0,2}^2}}}$\\
    & $+\frac{2\beta\left(1-\beta\right)}{\pi\left(r_{0,1}^2+r_{0,2}^2\right)}e^{-\frac{|\harpoon{v}|^2}{r_{0,1}^2+r_{0,2}^2}}$ & \\
     TH & $F_{pp}(\harpoon{v})=\frac{1}{\pi^2r_0^4}\left[2r_0^2\arccos{\left(\frac{|\harpoon{v}|}{2r_0}\right)}-\frac{v}{2}\sqrt{4r_0^2-v^2}\right]\theta(2r_0-|\harpoon{v}|)$ & $\frac{3\pi^3 r_0^2}{3\pi^2-16}$ \\
     \multirow{2}{*}{D}  & $F_{pp}(\harpoon{v})=\frac{m_g^4|\harpoon{v}|^2}{32\pi}K_0\left(m_g|\harpoon{v}|\right)+\frac{m_g^3|\harpoon{v}|\left(6+m_g^2|\harpoon{v}|^2\right)}{96\pi}K_1\left(m_g|\harpoon{v}|\right)$ & \multirow{2}{*}{ $\frac{28\pi}{m_g^2}$}\\
  & $+\frac{m_g^4|\harpoon{v}|^2}{96\pi}K_2\left(m_g|\harpoon{v}|\right)$ & \\
      \multirow{2}{*}{E} & $F_{pp}(\harpoon{v})=\frac{|\harpoon{v}|^2}{32\pi r_0^4}K_0\left(\frac{|\harpoon{v}|}{r_0}\right)+\frac{|\harpoon{v}|\left(6r_0^2+|\harpoon{v}|^2\right)}{96\pi r_0^5}K_1\left(\frac{|\harpoon{v}|}{r_0}\right)$ & \multirow{2}{*}{ $28\pi r_0^2$}\\
    & $+\frac{|\harpoon{v}|^2}{96\pi r_0^4}K_2\left(\frac{|\harpoon{v}|}{r_0}\right)$ & \\
     \hline
    \end{tabular}
    \caption{The analytical expressions of the overlap function $F_{pp}(\harpoon{v})$ and $\sigma_{{\rm eff},pp}$.}
    \label{tab:Fn}
\end{table*}

As an illustration, in the following, we take the $\langle \overrightarrow{r}^2 \rangle=(0.875~{\rm fm})^2$ for all profiles, where $0.875$ fm is the proton charge radius. Note that such values do not necessarily agree with other tunings. For instance, Ref.~\cite{Frankfurt:2003td} took $m_g^2=1.1$ GeV$^2$ from the analysis of the exclusive $J/\psi$ photoproduction (or electroproduction). Such a value results in the value of $\sqrt{\langle \overrightarrow{r}^2 \rangle}$ $1.5$ times smaller than $0.875$ fm. For the ``double Gaussian" profile, we adopt the values of $\beta=0.5,\frac{r_{0,1}}{r_{0,2}}=5$, as suggested in Ref.~\cite{Sjostrand:1987su}. In such a circumstance, we can predict the numerical values of $\sigma_{{\rm eff},pp}$ shown in the second column of Table~\ref{tab:sigmaeffvals}. The setup results in pretty large values of $\sigma_{{\rm eff},pp}$, ranging from $35$ mb with ``double Gaussian" to $70$ mb with ``top hat". Alternatively, we can also fix the value of $\sigma_{{\rm eff},pp}$ to extract the parameters. The predicted $\sqrt{\langle \overrightarrow{r}^2 \rangle}$ are displayed in the third column of Table~\ref{tab:sigmaeffvals} by using $\sigma_{{\rm eff},pp}=15$ mb. The $\sqrt{\langle \overrightarrow{r}^2 \rangle}$ values are generally $1.5-2.0$ times smaller than $0.875$ fm.

\begin{table*}[hbt!]
    \centering
    \begin{tabular}{|c|c|c|}
    \hline
      \multirow{2}{*}{Profile} & $\sigma_{{\rm eff},pp}$ (mb) & $\sqrt{\langle \overrightarrow{r}^2 \rangle}$ (fm)\\
       & ($\sqrt{\langle \overrightarrow{r}^2 \rangle}=0.875$ fm) & ($\sigma_{{\rm eff},pp}=15$ mb) \\
\hline
     HS  &   $69$  & $0.41$ \\
     G  &    $64$ &  $0.42$ \\
    DG &  $35$ & $0.58$ \\
     TH &  $70$ & $0.41$  \\
     D  &  $56$ & $0.45$ \\
     E & $56$ & $0.45$ \\
     \hline
    \end{tabular}
    \caption{The predictions of $\sigma_{{\rm eff},pp}$ after imposing the mean three-dimensional radius squared $0.875^2$ fm$^2$ (second column) and the values of the square root of the mean three-dimensional radius squared by fixing $\sigma_{{\rm eff},pp}=15$ mb (third column).}
    \label{tab:sigmaeffvals}
\end{table*}

\begin{figure}[hbt!]
\centering
\includegraphics[width=1.0\columnwidth,draft=false]{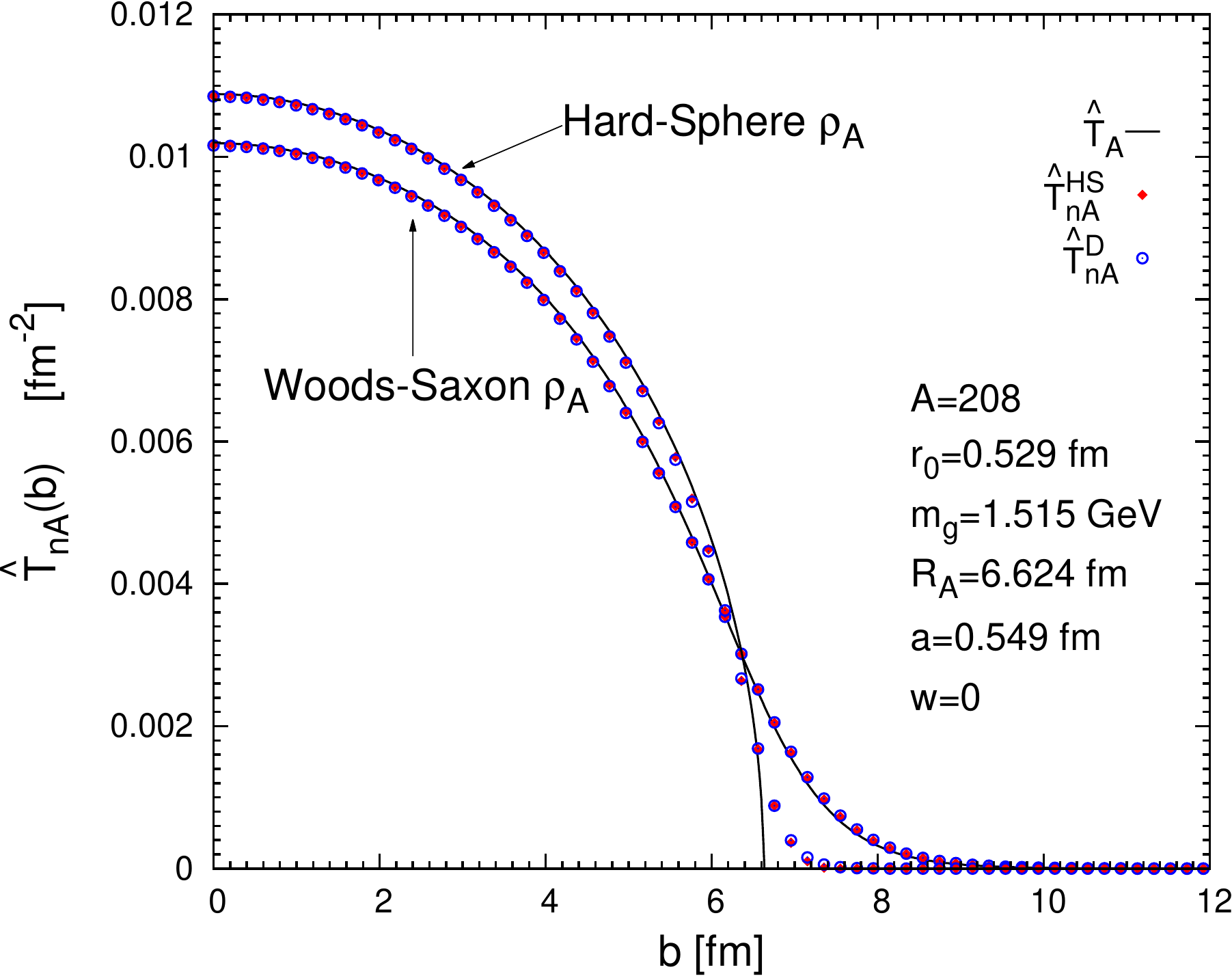}
\caption{\label{fig:TnAhat} The comparisons of $\hat{T}_{nA}(\harpoon{b})$ and $\hat{T}_{A}(\harpoon{b})$ for lead $A=208$.}
\end{figure}

In Fig.~\ref{fig:TnAhat}, we have shown the comparisons between $\hat{T}_{nA}(\harpoon{b})\equiv \int{F_{pp}(\harpoon{v})\hat{T}_A(\harpoon{v}-\harpoon{b})d^2\harpoon{v}}$ and $\hat{T}_{A}(\harpoon{b})$ for the lead $A=208$. Both the Woods-Saxon and hard-sphere $\rho_A$ have been used with the parameters $R_A=6.624$ fm, $a=0.549$ fm, and $w=0$. We have tried the two transverse parton profiles  HS and D. The approximation $\hat{T}_{nA}(\harpoon{b})\approx \hat{T}_{A}(\harpoon{b})$ is verified to be very good except where $b\equiv |\harpoon{b}|$ is close to the spherical surface ($b\simeq R_A$) in the hard-sphere $\rho_A$ case. Such a conclusion is quite general and should be independent of the functional form of the profile $t_p$ for nucleons in heavy nuclei. In particular, the consideration of the event-by-event fluctuation effect [cf., e.g., Eq.~(13) in Ref.~\cite{Mantysaari:2020axf}] in nucleons will not impact our results (bar the concrete value of $\sigma_{{\rm eff},pp}$). The details of the subnucelon structure, however, could be relevant in a description of light nuclei.

\section{The case of mutually different spatial distributions of proton and neutron inside nuclei\label{sec:pndiff}}

In this Appendix, we consider generalizing Eq.~(\ref{eq:mainAB}) for the case when the thickness functions for protons and neutrons in nuclei are different, e.g., because of the well-known neutron skin effect~\cite{Abrahamyan:2012gp,Tarbert:2013jze,Paukkunen:2015bwa} in nuclear physics. Such a generalization can be done by introducing the normalized proton and neutron thickness functions $\hat{T}_A^{p^A}(\harpoon{b})\equiv\hat{T}_A^{p}(\harpoon{b})$ and $\hat{T}_A^{n^A}(\harpoon{b})\equiv \hat{T}_A^{n}(\harpoon{b})$, while the thickness function is expressed as $T_A(\harpoon{b})=\sum_{N^A}{\hat{T}_A^{N^A}(\harpoon{b})}$. Similar to Eq.~(\ref{eq:TABdef}), we can introduce
\begin{eqnarray}
\hat{T}_{AB}^{N^AN^B}(\harpoon{b})&\equiv&\int_{-\infty}^{+\infty}{\hat{T}_A^{N^A}(\harpoon{s})\hat{T}_B^{N^B}(\harpoon{s}-\harpoon{b})d^2\harpoon{s}},
\end{eqnarray}
and we have
\begin{eqnarray}
T_{AB}(\harpoon{b})&=&\sum_{N^A,N^B}{\hat{T}_{AB}^{N^AN^B}(\harpoon{b})}. 
\end{eqnarray}
The GPDP in Eq.~(\ref{eq:gammaAij}) can be rewritten as
\begin{eqnarray}
&&\Gamma_A^{ij}(x_1,x_2,\harpoon{s}_1,\harpoon{s}_2,\harpoon{u}_1,\harpoon{u}_2)=\nonumber\\
&&\delta^2(\harpoon{s}_1-\harpoon{s}_2)\tilde{\Gamma}_{A}^{ij}(x_1,x_2,\harpoon{s}_1,\harpoon{u}_1,\harpoon{u}_2)\nonumber\\
&&+\frac{A-1}{2A}\left[\tilde{\Gamma}^i_{A}(x_1,\harpoon{s}_1,\harpoon{u}_1)\tilde{\Gamma}^j_{A}(x_2,\harpoon{s}_2,\harpoon{u}_2)\right.\nonumber\\
&&\left.+ \tilde{\Gamma}^i_{A}(x_1,\harpoon{s}_2,\harpoon{u}_1)\tilde{\Gamma}^j_{A}(x_2,\harpoon{s}_1,\harpoon{u}_2)\right],\label{eq:gammaAijnp}
\end{eqnarray}
with
\begin{eqnarray}
&&\tilde{\Gamma}^{ij}_{A}(x_1,x_2,\harpoon{s},\harpoon{u}_1,\harpoon{u}_2)\equiv\nonumber\\
&&\sum_{N^A}{\hat{T}^{N^A}_A(\harpoon{s})\Gamma^{ij}_{N^A}(x_1,x_2,\harpoon{s},\harpoon{u}_1,\harpoon{u}_2)}=\nonumber\\
&&Z\hat{T}^{p^A}_A\Gamma^{ij}_{p^A}+(A-Z)\hat{T}^{n^A}_A\Gamma^{ij}_{n^A},\nonumber\\
&&\tilde{\Gamma}^i_{A}(x,\harpoon{s},\harpoon{u})\equiv\nonumber\\
&&\sum_{N^A}{\hat{T}^{N^A}_A(\harpoon{s})\Gamma^{i}_{N^A}(x,\harpoon{s},\harpoon{u})}=\nonumber\\
&&Z\hat{T}^{p^A}_A\Gamma^{i}_{p^A}+(A-Z)\hat{T}^{n^A}_{A}\Gamma^{i}_{n^A}.
\end{eqnarray}
The transverse position function appearing in Eq.~(\ref{eq:bnPDF}) becomes $G\left(\frac{\hat{T}^{N^A}_A(\harpoon{s})}{\hat{T}^{N^A}_A(\harpoon{0})}\right)$, with the shorthand notations $G_{A,1}^{N^A}(\harpoon{s})\equiv G\left(\frac{\hat{T}^{N^A}_A(\harpoon{s})}{\hat{T}^{N^A}_A(\harpoon{0})}\right)$ and $G_{A,2}^{N^A}(\harpoon{s})\equiv 1-G_{A,1}^{N^A}(\harpoon{s})$. 

Then, Eq.~(\ref{eq:mainAB}) can be generalized by extending the indices $m_i$ in the transverse position symbols $\hat{T}$ to tuples $(m_i,N_i^A)$ when $i\in \{1,2\}$ and $(m_i,N_i^B)$ when $i\in \{3,4\}$. In other words, we have
\begin{eqnarray}
&&\sigma^{\rm DPS}_{AB\to f_1f_2}=\frac{1}{1+\delta_{f_1f_2}}\sum_{N_1^A,N_2^A,N_1^B,N_2^B}{\sum_{m_1,m_2,m_3,m_4=1}^{2}}\nonumber\\
&&\times \left(\sigma^{m_1m_3}_{N_1^AN_1^B\to f_1}\sigma^{m_2m_4}_{N_2^AN_2^B \to f_2}\right)\\
&&\times \left[\delta_{N_1^AN_2^A}\delta_{N_1^BN_2^B}\frac{\hat{T}_{A,(m_1,N_1^A)(m_2,N_2^A)}\hat{T}_{B,(m_3,N_3^B)(m_4,N_4^B)}}{\sigma_{{\rm eff},pp}}\right.\nonumber\\
&&+\delta_{N_1^BN_2^B}\frac{A-1}{A}\hat{T}^{(2)}_{nA,(m_1,N_1^A)(m_2,N_2^A)}\hat{T}_{B,(m_3,N_3^B)(m_4,N_4^B)}\nonumber\\
&&+\delta_{N_1^AN_2^A}\frac{B-1}{B}\hat{T}_{A,(m_1,N_1^A)(m_2,N_2^A)}\hat{T}^{(2)}_{nB,(m_3,N_3^B)(m_4,N_4^B)}\nonumber\\
&&\left.+\frac{(A-1)(B-1)}{AB}\hat{T}^{(2)}_{nAB,(m_1,N_1^A)(m_2,N_2^A)(m_3,N_3^B)(m_4,N_4^B)}\right].\label{eq:mainABpn}\nonumber
\end{eqnarray}
The new symbols are defined as
\begin{eqnarray}
&&\hat{T}_{A,(m_1,N_1^A)(m_2,N_2^A)}\equiv\int{\hat{T}_A^{N_1^A}(\harpoon{b})G_{A,m_1}^{N_1^A}(\harpoon{b})G_{A,m_2}^{N_2^A}(\harpoon{b})d^2\harpoon{b}},\nonumber\\
&&\hat{T}_{nA,(m_1,N_1^A)(m_2,N_2^A)}^{(2)}\equiv\nonumber\\
&&\int{\left(\hat{T}_{nA,(m_1,N_1^A)}(\harpoon{b})\hat{T}_{nA,(m_2,N_2^A)}(\harpoon{b})\right)d^2\harpoon{b}},\nonumber\\
&&\hat{T}^{(2)}_{nAB,(m_1,N_1^A)(m_2,N_2^A)(m_3,N_3^B)(m_4,N_4^B)}\equiv\nonumber\\
&&\frac{1}{2}\int{\left[\hat{T}_{nAB,(m_1,N_1^A)(m_3,N_3^B)}(\harpoon{b})\hat{T}_{nAB,(m_2,N_2^A)(m_4,N_4^B)}(\harpoon{b})\right.}\nonumber\\
&&\left.+\hat{T}_{nAB,(m_1,N_1^A)(m_4,N_4^B)}(\harpoon{b})\hat{T}_{nAB,(m_2,N_2^A)(m_3,N_3^B)}(\harpoon{b})\right]d^2\harpoon{b},\nonumber\\
&&\hat{T}_{nA,(m,N^A)}(\harpoon{b})\equiv\int{F_{pp}(\harpoon{v})\hat{T}_A^{N^A}(\harpoon{v}-\harpoon{b})G_{A,m}^{N^A}(\harpoon{v}-\harpoon{b})d^2\harpoon{v}},\nonumber\\
&&\hat{T}_{nAB,(m_1,N_1^A)(m_3,N_3^B)}(\harpoon{b})\equiv\nonumber\\
&&\int{F_{pp}(\harpoon{v})\hat{T}_{AB,(m_1,N_1^A)(m_3,N_3^B)}(\harpoon{b}-\harpoon{v})d^2\harpoon{v}},\nonumber\\
&&\hat{T}_{AB,(m_1,N_1^A)(m_3,N_3^B)}(\harpoon{b})\equiv\int{\left[\hat{T}_A^{N_1^A}(\harpoon{s})G_{A,m_1}^{N_1^A}(\harpoon{s})\right.}\nonumber\\
&&\left.\times\hat{T}_B^{N_3^B}(\harpoon{s}-\harpoon{b})G_{B,m_2}^{N_3^B}(\harpoon{s}-\harpoon{b})\right]d^2\harpoon{s}.
\end{eqnarray}
The $Ap$ counterpart (\ref{eq:mainAp}) can be generalized as
\begin{eqnarray}
&&\sigma^{\rm DPS}_{Ap\to f_1f_2}=\frac{1}{1+\delta_{f_1f_2}}\sum_{N_1^A,N_2^A}{\sum_{m_1,m_2=1}^{2}}\nonumber\\
&&\times \left(\sigma^{m_1 1}_{N_1^A p\to f_1}\sigma^{m_2 1}_{N_2^A p \to f_2}\right)\\
&&\times \left[\delta_{N_1^AN_2^A}\frac{\hat{T}_{A,(m_1,N_1^A)(m_2,N_2^A)}}{\sigma_{{\rm eff},pp}}+\frac{A-1}{A}\hat{T}^{(2)}_{nA,(m_1,N_1^A)(m_2,N_2^A)}\right].\label{eq:mainApnp}\nonumber
\end{eqnarray}

\bibliography{paper}

%
%
%
%
%

\end{document}